# The AR/VR Technology Stack: A Central Repository of Software Development Libraries, Platforms, and Tools

compiled by **Jasmine Roberts**[*]

**I. Low-Code Prototyping Tools**

**Adobe Aero** `adobe.com/products/aero.html`
Adobe Aero is an AR authoring tool that enables designers and creators to easily build interactive, immersive AR experiences using simple drag-and-drop tools. Aero empowers users to bring their 2D and 3D designs to life in AR by placing digital objects in real-world environments, allowing for new levels of engagement and interactivity. As a part of Adobe Creative Cloud, Aero integrates seamlessly with other Adobe tools and allows for easy exporting of AR projects to mobile devices. [1]

**Reality Composer** `apps.apple.com/us/app/reality-composer/id1462358802`
As a component of Apple's ARKit (see ARKit) framework, Reality Composer is a user-friendly visual editor that streamlines the creation of engaging and interactive AR content. It allows both technical and non-technical users to easily design and build immersive AR experiences by simply dragging and dropping 3D objects, animations, and other assets into a scene. [2]

**II. Ecosystem-Specific Creation Platforms**

**Effect House** `effecthouse.tiktok.com/`
TikTok's Effect House is a platform designed for creating and developing custom augmented reality effects for use on TikTok. The toolset offered on the platform includes features like face filters, 3D animations, and object recognition, which enable creators to develop engaging and interactive AR experiences. With TikTok Effect House, creators can share their AR effects with millions of TikTok users. [3]

**Lens Studio** `lensstudio.snapchat.com/`
Lens Studio is a comprehensive AR development tool created by Snap Inc. that enables creators to build engaging AR experiences for Snapchat users. It offers a range of features and tools that make it easy to design and develop custom lenses and filters, including real-time facial tracking, object recognition, and advanced animations. With Lens Studio, creators can leverage the power of AR to bring their ideas to life and share them with millions of Snapchat users around the world. [4]

**Spark AR** `sparkar.facebook.com/ar-studio/`
Spark AR is a platform for creating augmented reality experiences across Facebook, Instagram, and Messenger. With Spark AR, users can create custom filters, effects, and animations that can be shared with millions of users on social media platforms, making it an ideal tool for brands

[*]`jasminearvr@gmail.com`↗

and businesses looking to reach wider audiences with their content. [5]

**III. Web XR**

**8th Wall Reality Engine** `8thwall.com/tutorials`
The 8th Wall Reality Engine is an augmented reality (AR) platform acquired by Niantic Labs that enables developers to create high-quality AR experiences for mobile devices with ease. With its advanced computer vision technology and cloud-based AR rendering, the platform allows for markerless AR tracking and realistic 3D object placement. The 8th Wall Reality Engine also offers a range of features such as face effects, image recognition, and interactive animations. [6]

**A-Frame** `github.com/aframevr/aframe`
A-Frame is an open-source web framework developed by Josh Carpenter that provides developers with the tools to create immersive and interactive virtual reality experiences on the web. With A-Frame, developers can build VR experiences using HTML and JavaScript, making it accessible even to those without extensive knowledge of VR programming. The framework offers a range of features, including support for 3D models, animations, and physics engines, that make it easy to build rich and engaging VR experiences. It can also be used to create augmented reality experiences using web-based AR frameworks like AR.js. [7]

**AngularJS** `angularjs.org/`
AngularJS is a flexible and powerful tool for building web-based AR/VR applications. Its modular architecture and built-in features make it easier for developers to create complex applications, and its popularity within the web development community ensures a wealth of resources and support. [8]

**AR.js** `ar-js-org.github.io/AR.js-Docs/`
AR.js is an open-source JavaScript library designed for building web-based augmented reality experiences. It uses computer vision technology to track markers and images in real time, allowing developers to create AR applications that run on mobile devices and desktop browsers without the need for additional software or plugins. AR.js provides developers with a range of features, including support for 3D models, animations, and sound, as well as the ability to add interactivity to AR scenes. [9]

**BabylonJS** `www.babylonjs.com/`
BabylonJS is an open-source JavaScript framework for building 3D games and applications that can run in web browsers. It offers a range of tools and features such as WebGL, WebVR, physics engines, and animations. With BabylonJS, developers can create rich and immersive 3D environments that can be accessed on desktops and mobile devices through web browsers. It is widely used and supported by a community of contributors on GitHub. [10]

**JanusWeb** `github.com/jbaicoianu/janusweb`

JanusWeb is a browser-based, open-source project that enables developers to create and share immersive virtual reality (VR) experiences. It allows developers to design and deploy interactive 3D scenes that can be accessed on any modern web browser on desktops and mobile devices. With its easy-to-use interface and a host of features such as 3D graphics, audio, and scripting, JanusWeb is an ideal platform for building and sharing VR experiences on the web. It is also backed by an active community of contributors on GitHub [11]

**JavaScript** `javascript.com/`
JavaScript is a programming language that can be used in combination with most of the frameworks and tools that are previously mentioned, like A-Frame, BabylonJS, and AR.js. JavaScript can be used to create and control the behavior of web-based AR/VR experiences, such as interactive 3D scenes, animations, and user interfaces. It can also be used to integrate with other web technologies, like WebXR and WebVR, which provide support for immersive experiences on the web. [12]

**PrimroseVR** `primrosevr.com/`
Primrose VR is an open-source JavaScript framework built on top of A-Frame and Three.js that enables developers to create immersive virtual reality experiences for the web. It provides a range of features, including support for stereoscopic rendering, spatial audio, motion tracking, and more. With Primrose VR, developers can easily create and deploy VR experiences that can be accessed using any web browser or VR headset. [13]

**React VR** `npmjs.com/package/react-vr`
React VR is a library for building AR/VR applications using ReactJS. It provides a range of tools and features for easy creation of immersive experiences that run in web browsers and on mobile devices. With React VR, developers can create interactive user interfaces, manage 3D objects and animations, and connect to AR/VR hardware and sensors. This library is well suited for creating VR applications that are cross-platform compatible, scalable, and easy to maintain. [14]

**three.JS** `threejs.org`
ThreeJS is a free and open-source JavaScript library developed and maintained by Ricardo Cabello for creating 3D computer graphics in web browsers. It provides a range of powerful tools and features, including support for WebGL, shaders, lighting, cameras, animations, and more. With ThreeJS, developers can create 3D visualizations and games that run in web browsers on desktops and mobile devices. [15]

**IV. General SDKs**

**ARCore** `developers.google.com/ar/`
ARCore is a software development kit (SDK) created by Google that enables developers to build augmented reality (AR) applications for Android devices. With ARCore, developers can use the device's camera to place virtual objects in the real world and create immersive AR experiences. The SDK provides access to a range of AR features, including motion tracking, environmental understanding, and light estimation, which help developers create realistic and compelling AR content. [16]

**ARKit** `developer.apple.com/augmented-reality/`
ARKit is a software development kit (SDK) created by Apple for building augmented reality (AR) applications on iOS devices. It provides developers with the tools to create immersive AR experiences by allowing them to access the device's camera and motion sensors to track the user's position and movements, as well as to detect real-world objects and surfaces. ARKit offers a range of features, including 3D object recognition, lighting estimation, and surface detection. [17]

**CameraKit SDK** `kit.snapchat.com/camera-kit`
Camera Kit is a cross-platform software development kit designed by Snap Inc. to provide a suite of tools and capabilities for building augmented reality experiences that can be accessed through Snapchat's camera. It enables developers to create AR lenses, filters, and effects using a variety of programming languages, including C++, Objective-C, and Java. The SDK includes a range of features such as 3D object tracking, facial tracking, and image recognition, and can be integrated with popular game engines such as Unity and Unreal Engine. [18]

**Cloud XR SDK** `developer.nvidia.com/nvidia-cloudxr-sdk`
Cloud XR SDK is a tool by NVIDIA for creating AR and VR experiences that allows developers to stream high-fidelity 3D content from the cloud to any device. The SDK offers cross-platform compatibility with major AR and VR devices, including Oculus Quest, HoloLens 2, and Android devices, among others. It also offers features such as real-time ray tracing, 360-degree video, and low-latency streaming [19]

**Google VR SDK** *(now Cardboard SDK)*
`developers.google.com/vr/develop/unity/get-started-android`
Cardboard was first introduced in 2014 and is designed to be a low-cost, accessible way for people to experience virtual reality. Cardboard SDK is a set of tools for developers to build virtual reality experiences for the platform. The SDK is compatible with both Android and iOS devices, and it includes features such as head tracking, spatial audio, and gesture recognition. [20]

**Lightship ARDK** `lightship.dev/`
Lightship ARDK, developed by Niantic, is a comprehensive augmented reality development platform designed to build and deploy AR experiences for mobile devices. It includes a range of features such as real-time mapping, occlusion, surface detection, and light estimation, making it possible to create more immersive and realistic AR experiences. With the Lightship ARDK, developers can easily create AR experiences for both Android and iOS devices and take advantage of Niantic's extensive location-based technology expertise. [21]

**Lumin SDK**
`https://developer.magicleap.com/downloads/lumin-sdk/`
Lumin SDK is a development platform for creating spatial computing experiences for Lumin OS, a proprietary operating system developed by Magic Leap. It provides developers with tools to create augmented reality applications that are responsive to real-world environments, enabling users to interact with digital content as if it were physically present in the world around them. The Lumin SDK provides access to a range of sensors, including eye tracking, gesture recognition, and environmental tracking, which allow developers to create highly immersive experiences that respond to a user's natural movements and interactions. Lumin SDK also supports Unity, Unreal Engine, and WebGL. [22]

**Mixed Reality Extension SDK**
`github.com/Microsoft/mixed-reality-extension-sdk`
The Mixed Reality Extension SDK, developed by Microsoft, is a powerful tool for building immersive experiences that merge the physical and digital worlds. It provides a platform for developers to create cross-platform mixed reality apps that can run on various devices and platforms. With its advanced features, the Mixed Reality Extension SDK allows developers to build complex scenarios, create realistic 3D models, and interact with the physical environment using digital objects. It also offers an easy-to-use authoring tool that simplifies the development process and allows developers to focus on building engaging and immersive experiences. [23]

**MRTK** `docs.microsoft.com/en-us/windows/mixed-reality/mrtk-unity/`
The Mixed Reality Toolkit (MRTK) is an open-source, cross-platform development kit for building mixed reality applications. It offers a set of components and features that help developers create immersive experiences across a variety of platforms, including Microsoft HoloLens, Windows Mixed Reality headsets, and Oculus Quest. MRTK provides a variety of tools for building applications, such as hand and eye-tracking, spatial mapping, and voice recognition. It is designed to be extensible, so developers can customize and extend the toolkit to fit their specific needs. [24]

**NRSDK** `developer.nreal.ai/download`
NRSDK is a platform developed by Nreal for creating mixed reality experiences. It offers a high-level API and simple development process, enabling Nreal glasses to understand the real world through spatial computing, optimized rendering, and multi-modal interactions. NRSDK's spatial computing features include 6DoF tracking, plane detection, image tracking, and hand tracking. It accurately tracks the position and orientation of the glasses as they move through space and recognizes physical images and hand poses in real-time. The optimized rendering performance in the backend reduces latency and judder, providing a smooth and comfortable user experience. Multi-modal interactions are also provided, including Nreal Light Controller (3DoF) and Nreal Phone Controller (3DoF), which enable natural and intuitive ways to interact with virtual objects. [25]

**Oculus SDK** `developer.oculus.com/`
The Oculus SDK (Software Development Kit) is a set of tools and resources that developers can use to create applications and games for Oculus virtual reality devices. It includes a wide range of features, such as tracking, rendering, and user input, as well as advanced capabilities like spatial audio, hand tracking, and social features. The SDK supports a variety of platforms and programming languages, and provides extensive documentation and support to help developers get started and succeed in their projects. [26]

**OpenFrameworks** `openframeworks.cc/`
OpenFrameworks is a creative coding toolkit and development framework developed by Zachary Lieberman for building interactive and multimedia applications. It is an open-source, cross-platform toolkit that provides a wide range of libraries, tools, and plugins for creating applications in C++. OpenFrameworks can be used for generative art, interactive installations, data visualization, computer vision, and virtual and augmented reality experiences. In the context of AR and VR, OpenFrameworks can be used with ARKit and ARCore to create immersive and interactive experiences that blend digital content with the physical world. [27]

**OpenVR SDK** `github.com/ValveSoftware/openvr`
OpenVR is an open-source virtual reality software development kit (SDK) that enables developers to create VR experiences that can run on different VR headsets. It was developed by Valve Corporation and supports various VR devices, such as HTC Vive, Oculus Rift, and Windows Mixed Reality. With OpenVR, developers can access VR system information, including the position and orientation of the user's head, controllers, and other devices. They can also develop applications that can work with multiple VR devices, making it an ideal choice for cross-platform VR development. Additionally, OpenVR provides a wide range of tools and features, such as VR rendering, tracking, and input management. [28]

**Pico Unity Integration SDK** `developer-global.pico-interactive.com/sdk`
Pico Unity Integration SDK, developed by Bytedance's Pico Interactive, offers various features for VR app development. It supports rendering features such as late latching and space warp, and also includes eye and face tracking for input and tracking. The SDK also provides developer tools like Metrics HUD and PICO Developer Center, and added APIs for customizing the app library and retrieving controller information. The SDK supports IAP, Interaction & Challenge, Leaderboard, Achievements, and debugging RTC and Challenge services. The latest version also includes support for Vulkan and haptic feedback. [29]

**Reality Kit** `developer.apple.com/documentation/realitykit`
Reality Kit is a framework developed by Apple for creating high-quality augmented reality experiences on iOS

devices. It provides developers with a range of tools and features for building AR apps, including support for 3D models, animations, physics simulations, and spatial audio. Reality Kit also includes a feature called Reality Composer (see Reality Composer), which allows developers to visually create interactive AR scenes without writing any code. With Reality Kit, developers can create immersive AR experiences that blend the physical and digital worlds together in new and exciting ways. [30]

**Snapdragon Spaces**
`qualcomm.com/products/features/snapdragon-spaces-xr-platform/`
Snapdragon Spaces is a spatial computing platform developed by Qualcomm Technologies. It allows developers to create immersive and interactive experiences on standalone AR and VR devices, using features like room mapping, object recognition, and hand tracking. The platform also supports cloud-based services, making it easier to distribute and manage content across multiple devices [31]

**Tobii XR SDK** `vr.tobii.com/sdk/`
Tobii XR SDK is a cross-platform software development kit that enables the integration of eye tracking and head tracking into virtual and augmented reality applications. The SDK provides developers with access to Tobii's advanced eye tracking technology, which allows for more immersive and intuitive user experiences. It also includes tools for visualizing and analyzing eye tracking data, as well as support for a wide range of platforms and engines, including Unity and Unreal Engine. [32]

**Vive Sense SDK**
`developer.vive.com/resources/vive-sense/`
The Vive Sense SDK is a software development kit that allows developers to integrate the capabilities of the Vive Pro Eye headset and other compatible devices, such as the Vive Focus Plus, into their applications. The SDK provides access to eye tracking data, hand tracking, and object recognition. The Vive Sense SDK includes support for both Unity and Unreal Engine. [33]

**VRTK** `vrtk.io/`
VRTK is a popular open-source toolkit for developing VR applications on Unity. It provides a collection of useful scripts and components for building VR interactions, such as teleportation, grabbing and throwing objects, and locomotion. VRTK also supports a wide range of VR devices, including Oculus Rift, HTC Vive, Windows Mixed Reality, and more. Its modular design allows developers to customize and extend it according to their specific needs, making it a versatile and powerful tool for creating immersive VR experiences. Additionally, VRTK has an active community of developers who provide support and share their own contributions, making it a collaborative and constantly evolving project. [34]

**Vuforia** `developer.vuforia.com/downloads/sdk`
Vuforia is an augmented reality (AR) software development kit (SDK) developed by PTC Inc. It allows developers to create AR applications that can recognize and track images and objects in the real world, and overlay digital content on top of them in real time. Vuforia supports multiple platforms including iOS, Android, and Unity, and has been used in a wide range of AR applications, from marketing and advertising to gaming and education. Its features include image recognition and tracking, 3D object recognition and tracking, text recognition, and more. [35]

**Wikitude** `https://www.wikitude.com/products/wikitude-sdk/`
Wikitude is an augmented reality SDK that enables developers to create AR apps for both mobile and smart glasses. The platform offers a wide range of features, including image and object recognition, geo-location AR, and markerless SLAM technology. It also supports various programming languages and platforms, such as JavaScript, Unity, Xamarin, and Cordova. Wikitude also offers a cloud recognition service that allows developers to create and manage their own image recognition databases. [36]

**XRTK** `xrtk.io/`
The Mixed Reality Toolkit (XRTK) is an open-source framework for developing mixed reality applications, providing developers with a set of reusable components and tools for building AR/VR experiences across different platforms. With XRTK, developers can focus on building immersive user experiences without worrying about the underlying platform-specific technical details. The toolkit supports multiple platforms, including HoloLens, Oculus, and Windows Mixed Reality, and provides out-of-the-box support for features such as input management, spatial mapping, and cross-platform networking. [37]

**V. Multiplayer SDKs**

**Mirror** `https://mirror-networking.com/`
Mirror is a high-performance networking library for Unity game engine that simplifies the process of creating multiplayer games and applications for AR and VR. With Mirror, developers can build multiplayer experiences that can be run on a variety of platforms, including desktops, mobile devices, and VR headsets. The library provides tools for network communication, object synchronization, and player management, allowing developers to create complex and immersive multiplayer environments with ease. [38]

**NetCode for GameObjects** `docs-multiplayer.unity3d.com/netcode/current/about/index.html`
Netcode for GameObjects is a Unity package that offers networking capabilities to GameObject & MonoBehaviour workflows. It is interoperable with multiple low-level transports, including the official Unity Transport Package. This high-level networking library is designed for Unity to abstract networking logic. Netcode enables you to send GameObjects and world data across a networking session to many players at once, allowing you to focus on game development instead of dealing with low-level protocols and networking frameworks. [39]

**Normcore** `normcore.io/`
Normcore is a multiplayer platform for creating social VR experiences. It allows developers to easily create and customize avatars, as well as to implement networking features such as voice chat and synchronized interactions. Normcore is designed to be compatible with a variety of VR and AR platforms, including Oculus, Vive, and HoloLens. Its goal is to make it easier for developers to create immersive, social experiences in VR and AR. [40]

**Photon Fusion**
`doc.photonengine.com/fusion/current/getting-started/fusion-intro`
Fusion is a new networking library built for Unity that offers high performance state synchronization. It provides a single API to support two different network topologies and a single player mode with no network connection. The library is designed to integrate seamlessly into Unity workflow while also offering advanced features such as data compression, client-side prediction, and lag compensation. Network objects can be defined as prefabs and network state is defined using attributes on the MonoBehaviour itself. Fusion relies on a compression algorithm to reduce bandwidth requirements and uses tick-based simulation to ensure robust network performance. It operates in either Shared Mode or Hosted Mode, depending on who has authority over network objects. [41]

**Photon PUN** `photonengine.com/sdks`
Photon PUN (Photon Unity Networking) is a plugin for Unity that makes it easy to add multiplayer functionality to AR/VR applications. It provides a simple and scalable solution for synchronizing game state and player actions across multiple devices and platforms.  No longer supported in favor of Photon Fusion. [42]

**Photon Voice** `https://www.photounengine.com/sdks`
Photon Voice is a real-time voice and audio communication solution that allows developers to easily add voice chat to their multiplayer games and applications. With Photon Voice, developers can create rich and immersive audio experiences for their users by incorporating features such as echo cancellation, noise reduction, and voice activity detection. This solution is built on top of Photon's high-performance networking engine, which ensures reliable and low-latency audio communication. [43]

**Shared AR** *(see Lightship ARDK)*

**VI. APIs**

**ARCore API** `arvr.google.com/arcore/`
ARCore is actually an API developed by Google for creating AR experiences on Android devices. It provides tools for motion tracking, environmental understanding, and light estimation, allowing developers to create immersive AR experiences that integrate seamlessly with the user's physical surroundings. With ARCore, developers can build AR apps that can detect flat surfaces, place virtual objects in real-world environments, and enable interactive experiences through touch and motion controls. It is compatible with a range of Android devices, and is equipped with features like Cloud Anchors and Sceneform support. [44]

**OpenCV** `https://opencv.org/`
OpenCV (Open Source Computer Vision Library) is primarily an open-source computer vision and machine learning software library, which provides a collection of algorithms and tools for image processing, object detection and recognition, video analysis, and more. It can be used as an API for developing AR and VR applications in a variety of programming languages, including C++, Python, and Java, and it can be combined with other AR and VR engines, SDKs, and platforms to create more complex applications. [45]

**Open XR** `https://www.khronos.org/OpenXR/`
OpenXR is an open standard API for developing VR and AR applications that allows developers to create applications that work across multiple hardware platforms. It provides a standardized interface for interacting with VR and AR devices, allowing developers to create applications that can be used on different platforms without requiring them to write separate code for each platform. OpenXR is designed to simplify the development process, making it easier for developers to create high-quality VR and AR applications that work across a range of devices. The standard is maintained by the Khronos Group, a non-profit consortium of companies focused on creating open standards for graphics and media. [46]

**WebAudio** `developer.mozilla.org/en-US/docs/Web/API/Web_Audio_API`
Web Audio is an API that allows developers to create and manipulate audio content on the web. It provides a set of JavaScript interfaces for processing and synthesizing audio in real-time, allowing for a variety of audio effects to be applied to soundscapes, music, and voice. With Web Audio, developers can create immersive sound experiences for AR/VR applications, interactive music applications, and web-based games. This API also supports spatial audio. [47]

**WebXR** `developer.mozilla.org/en-US/docs/Web/API/WebXR_Device_API/Fundamentals`
Mozilla WebXR is an open standard and API that enables developers to create augmented reality (AR) and virtual reality (VR) experiences that run on the web, across a wide range of devices and platforms. With WebXR, developers can create immersive experiences that can be accessed using web browsers and headsets, making it easier for users to access AR/VR content without having to download or install any additional software. The WebXR API supports a range of input devices, including gamepads, touchscreens, and motion controllers, allowing developers to create interactive and responsive experiences. WebXR also provides a variety of features such as spatial audio and hand tracking. [48]

## VII. Protocols

**GraphQL** `graphql.org/`
GraphQL is a query language and runtime that allows developers to define the structure of data and operations that can be performed on that data. It is often used in AR/VR applications to provide a more efficient and flexible way of communicating between the client and the server. With GraphQL, developers can specify exactly what data they need, and the server will only return that data, reducing the amount of unnecessary data transfer and improving performance. GraphQL also provides tools for real-time updates, making it an ideal choice for building interactive AR/VR experiences that require fast and dynamic data exchange. [49]

**REST**
REST, or Representational State Transfer, is an architectural style that defines a set of constraints for creating web services often used in AR/VR applications to provide a standardized way of communicating between the client and the server. By using RESTful APIs, developers can easily retrieve, manipulate and store data, making it an essential technology for building complex AR/VR experiences. REST provides a scalable and efficient way of creating web services that can handle large amounts of data and requests. [50]

**SOAP** `www.w3.org/TR/soap/`
(Simple Object Access Protocol) is a messaging protocol used for exchanging structured information between web services. SOAP could be used to send data between an AR/VR application and a web service that provides information on real-world objects or locations. SOAP can be used to implement APIs for AR/VR systems, allowing developers to expose functionality and data to other applications and services. However, other web service protocols, such as REST and GraphQL may be better suited for AR/VR applications. [51]

**WebRTC** `webrtc.org/`
WebRTC is an open-source technology that enables real-time communication over the internet, such as voice and video chat, without the need for plugins or external software. In the context of AR/VR, WebRTC is commonly used to provide low-latency, high-quality audio and video communication between multiple users within virtual environments. This can include enabling remote collaboration between users within a shared virtual workspace or facilitating multiplayer gaming experiences within VR environments. [52]

**WebSockets** `developer.mozilla.org/en-US/docs/Web/API/WebSockets_API`
WebSockets are a protocol that provides a persistent, bidirectional communication channel between a client and a server. In the context of AR/VR, WebSockets can be used to establish a real-time connection between the user's device and a server to exchange data related to the AR/VR experience. For example, it can be used to synchronize the position and orientation of objects in a shared AR environment, or to stream real-time video data from a VR headset to a remote server for processing. WebSockets provide low-latency communication and can be used in conjunction with other technologies such as WebGL, WebRTC, and WebVR to create immersive and interactive AR/VR experiences. [53]

**VIII. Databasing**

**MySQL** `dev.mysql.com/doc/`
MySQL is an open-source relational database management system that can be used to store and manage large amounts of data. With MySQL, developers can create, read, update, and delete data related to user accounts, 3D models, and other relevant data. MySQL also supports features such as replication, clustering, and backups. Furthermore, MySQL can be used in combination with other technologies, like Node.js and Unity, to create powerful and scalable AR/VR solutions. MySQL can be used to store user information and preferences in a Node.js server, which can then be retrieved and used in a Unity-based AR/VR application. [54]

**NoSQL** `docs.oracle.com/en/database/other-databases/nosql-database/`
NoSQL is a type of database management system that is well-suited for handling large volumes of unstructured or semi-structured data. NoSQL databases can be used to store and manage the massive amounts of data generated by AR/VR applications *e.g.* user interactions, sensor data, and object recognition results. NoSQL databases are often used in conjunction with real-time data processing and analytics tools to enable faster insights and decision-making in applications. Popular NoSQL databases used in AR/VR development include MongoDB, Cassandra, and CouchBase. [55]

**PostgreSQL** `postgresql.org/docs/`
PostgreSQL is an open-source relational database management system that is often used as a data storage solution. With its support for advanced features such as JSON and spatial data, PostgreSQL can help to enhance the performance of AR/VR applications by enabling them to efficiently store and retrieve complex data structures. Additionally, PostgreSQL's support for ACID transactions ensures data consistency and reliability, which is particularly important for real-time applications such as those found in the AR/VR space. [56]

**IX. Common File Formats**

**fbx** `autodesk.com/products/fbx/overview`
FBX is a file format developed by Autodesk for 3D content creation, including modeling, animation, and rendering. It is widely used in the gaming and entertainment industries, and is compatible with many software applications as such FBX is widely considered one of the most interoperable file formats. [57]

**GLB**
GLB is a file format used to represent 3D models with textures and animations. It is the binary file format representation of gltf. [58]

**glTF** `khronos.org/gltf/`
glTF is an open standard file format for 3D assets, developed by the Khronos Group. It was created to address the need for a common, efficient, and interoperable format for transmitting 3D models, textures, and other assets across different authoring tools, services, and platforms. The glTF format uses a JSON-based structure that describes the geometry, materials, animations, and other properties of a 3D object, which can be compressed to reduce file size and transmitted over the web using standard protocols like HTTP or WebSocket. glTF supports a range of features, including PBR (physically-based rendering) materials, skeletal animations, morph targets, instancing, and more. It is widely supported by 3D engines, applications, and frameworks. [59]

**OBJ**
OBJ is known for its ability to store detailed geometric information, such as vertices, edges, and faces, as well as material and texture data, making it an ideal format for creating realistic and intricate 3D models. [60]

**USDZ**
`developer.apple.com/documentation/arkit/usdz_schemas_for_ar`
USDZ is a file format developed by Apple and Pixar for AR/VR applications. It is used to represent 3D models with textures, animations, and other features, and is optimized for use in Apple's ARKit framework. [61]

**VRML** `w3.org/MarkUp/VRML/`
VRML stands for Virtual Reality Modeling Language, and it is a 3D graphics file format used for displaying interactive 3D content, including virtual reality and augmented reality applications. It was developed in the mid-1990s as a way to create 3D virtual worlds that could be viewed on the web, and it has since been used in a variety of AR and VR applications. VRML provides a way to create complex 3D models that can be rendered in real time, allowing users to interact with the virtual environment in a way that feels immersive and interactive. While VRML is not as widely used as some other 3D file formats, it remains an important part of the history of AR and VR technology, and it continues to be used in some specialized applications today. [62]

**X3D** `https://www.web3d.org/x3d/what-x3d`
X3D is an open standard for 3D graphics on the web, designed to support a wide range of applications, including augmented reality and virtual reality. It is a successor to VRML (Virtual Reality Modeling Language) and provides improved support for advanced features such as lighting, shaders, and animations. X3D files can be viewed in web browsers that support the X3D standard, or in specialized viewers that can be integrated into AR and VR applications. [63]

**xVRML** `sourceforge.net/projects/xvrml/`
xVRML is an extension of VRML (Virtual Reality Modeling Language) that includes support for Extensible Markup Language (XML). Like VRML, xVRML is a file format used for describing 3D scenes and interactive objects for web-based applications, including those in the AR/VR space. xVRML extends the capabilities of VRML by providing a more robust set of features for representing complex 3D objects, animations, and interactions. [64]

## X. Backend

**C#** `docs.microsoft.com/en-us/dotnet/csharp/`
C# is a widely-used programming language, notably for creating immersive experiences with the Unity game engine. Its powerful features, such as garbage collection, strong typing, and asynchronous programming, make it a popular choice for building complex and interactive applications that run smoothly on various platforms. [65]

**C++** `cplusplus.com/doc/tutorial/`
C++ is a programming language that has been used extensively in AR/VR development. Its speed, efficiency, and ability to handle complex computations make it an attractive choice for building high-performance AR/VR experiences. Many AR/VR development tools and engines, including Unreal Engine and Unity, use C++ as their primary programming language. [66]

**Java** `docs.oracle.com/en/java/`
Java is the most used object-oriented programming language and has ability to run on various platforms, including desktops, mobile devices, and game consoles. Java has become an essential tool for developers looking to create immersive experiences across different devices. Additionally, the language's robust libraries, such as Java3D and jMonkeyEngine, provide developers with powerful tools for building and rendering 3D graphics, physics engines, and other advanced features. [67]

**Lua** `lua.org/docs.html`
Lua is a lightweight and fast scripting language commonly used in game development and other performance-sensitive applications. Its ease of use and flexibility make it an ideal choice for AR/VR development, especially for creating interactive and dynamic content. Roblox uses Lua as its primary scripting language for game development. [68]

**.NET** `docs.microsoft.com/en-us/dotnet/`
.NET is a versatile programming framework primarily used for Windows desktop and web application development. .NET can be used to build mixed reality applications for Windows Mixed Reality headsets, and to develop backend services for AR/VR applications hosted on Azure. [69]

**Node.js** `nodejs.org/en/docs/`
Node.js is a popular JavaScript runtime built on Chrome's V8 JavaScript engine. It allows developers to use

JavaScript on the server-side to build scalable and fast web applications. Node.js is used in AR/VR development to build server-side applications that interact with AR/VR devices and provide real-time data to these devices. [70]

**Objective-C**
`https://developer.apple.com/library/archive/documentation/Cocoa/Conceptual/ProgrammingWithObjectiveC/Introduction/Introduction.html#//apple_ref/doc/uid/TP40011210`
Objective-C is a high-level, object-oriented programming language used to develop applications for Apple's platforms, including iOS, macOS, and watchOS. With dynamic binding and messaging, it is well-suited for developing AR/VR applications that require complex interactions between objects and users. Objective-C is also the primary language used to develop ARKit, Apple's augmented reality platform, making it an essential tool for developers looking to build immersive AR experiences. [71]

**Python** `docs.python.org/3/`
Python is a high-level, interpreted programming language that has become increasingly popular in AR/VR development due to its ease of use, large community support, and the availability of powerful libraries like NumPy, OpenCV, and Pygame. It can be used to develop AR/VR applications that incorporate computer vision, machine learning, and other advanced technologies. Python's versatility and cross-platform compatibility make it a valuable tool for developing AR/VR applications that can run on a wide range of devices and platforms. [72]

**Ruby** `ruby-lang.org/en/documentation/`
Ruby is a dynamic, high-level programming language that has been used in AR/VR development to build applications and games. With Ruby, developers can build AR/VR applications that run on different platforms, including mobile devices and desktops. [73]

**Swift** `swift.org/documentation/`
Swift is a powerful and intuitive programming language developed by Apple for building apps for iOS, macOS, watchOS, and tvOS. It is used extensively for developing ARKit-based AR applications and has rapidly gained popularity among developers due to its ability to create high-performance, native apps that can run smoothly on Apple devices. Swift has a robust library of tools and support for real-time rendering and animation. [74]

**XI. Engines & Platforms**

**Amazon Sumerian** `aws.amazon.com/sumerian/`
Amazon Sumerian is a browser-based development platform and set of tools for building and deploying augmented reality, virtual reality, and 3D applications. It allows developers to create immersive experiences using pre-built objects and customizable templates, as well as import 3D assets from third-party tools. Additionally, Sumerian provides integration with other Amazon Web Services, such as Amazon Polly for speech and Amazon Lex for conversational interfaces. [75]

**Android** `developer.android.com/`

Android is a mobile operating system developed by Google. It provides developers with a powerful platform for building AR/VR applications through its support for high-performance graphics, 3D rendering, and sensor integration. Additionally, Android provides APIs for accessing device features such as cameras, microphones, and GPS. As an open-source platform, it is highly customizable and has a large community of developers constantly creating new tools and frameworks to enhance AR/VR experiences. The Oculus Quest, a popular standalone VR headset, is powered by a modified version of Android, allowing developers to leverage their existing knowledge of the platform to create immersive VR experiences for the device. [76]

**Azure** `azure.microsoft.com/`
Azure is a cloud computing platform and service provided by Microsoft that offers a range of tools and services for building and deploying various applications. One of the key benefits of Azure is its scalability, as it allows developers to easily scale up or down their applications depending on demand. Azure also provides a range of tools for data storage, analytics, and machine learning, which can be useful in developing AR/VR applications that require complex data processing. Additionally, Azure provides various tools for building and deploying web-based experiences, including support for 3D graphics, audio, and video. [77]

**EasyAR** `easyar.com/`
EasyAR is an augmented reality engine that provides developers with a comprehensive set of tools to create AR applications for mobile devices. It supports a variety of platforms including iOS, Android, and Unity, and provides features such as image tracking, face tracking, and 3D object tracking. With EasyAR, developers can easily create engaging AR experiences that are both interactive and immersive. The engine also provides a user-friendly interface, making it accessible for developers of all skill levels. [78]

**Godot** `github.com/GodotVR/`
The Godot VR Engine is an open-source game development engine designed specifically for building immersive virtual reality experiences. This powerful engine offers developers a flexible and feature-rich toolset to design, prototype, and deploy VR applications across various platforms. Godot VR Engine supports a range of VR devices and offers a variety of tools such as physics simulation, audio, and scripting to enable developers to create engaging and interactive VR content. The engine also boasts a supportive community and a comprehensive documentation library to help developers navigate its features and capabilities. It's important to note that the quality of the VR experiences created using Godot VR Engine depends on a variety of factors, including the creativity and skills of the developer. [79]

**iOS** `developer.apple.com/ios/`
iOS is a mobile operating system developed by Apple Inc. that is used in its iPhone and iPad devices. It is known for its user-friendly interface, high security standards, and its integration with Apple's proprietary software and services such as Siri, iCloud, and Apple Pay. In the context of AR/VR, iOS has become an important platform for developers due to its large user base and the availability of ARKit, Apple's augmented reality development platform. ARKit allows developers to create immersive AR experiences that can be integrated into iOS apps. Additionally, iOS devices such as the iPhone and iPad are often used as controllers for VR experiences, making iOS a key player in the AR/VR space. [80]

**Omniverse** `developer.nvidia.com/nvidia-omniverse-platform/`
NVIDIA Omniverse is a virtual collaboration platform that enables teams to work together on complex 3D projects in real-time. It allows for the creation, simulation, and rendering of photorealistic environments and objects, making it a powerful tool for AR/VR development. With its open architecture, Omniverse can integrate with a wide range of software tools and platforms, such as Unity, Unreal Engine, and Autodesk. This allows for seamless workflows and collaboration across different teams, while also providing access to a variety of assets and resources. Additionally, Omniverse features AI-powered physics simulation and rendering capabilities. [81]

**PlayCanvas** `playcanvas.com/`
PlayCanvas is an open-source game engine designed for creating interactive 3D experiences in the browser. It uses WebGL and HTML5 technologies, making it a popular choice for web-based VR and AR experiences. PlayCanvas has a user-friendly interface and supports real-time collaboration, allowing developers to work together on a project remotely. It also offers a built-in physics engine, asset management, and scripting tools to create complex interactive experiences. PlayCanvas can export to multiple platforms, including VR headsets, mobile devices, and desktop browsers. [82]

**Roblox** `roblox.com/create/`
Roblox is a popular online game creation platform that allows users to create and play their own games. Roblox has added support for VR, allowing players to experience games and virtual worlds in a fully immersive way. The platform also offers support for AR, enabling developers to create experiences that blend the physical world with digital elements. The Roblox Developer Hub provides a suite of tools and resources for creating immersive AR/VR experiences, including a robust scripting language, asset creation tools, and an extensive library of community-created assets. [83]

**Source 2** `developer.valvesoftware.com/wiki/Source_2`
The Source 2 game engine is a powerful game engine developed by Valve Corporation, which has been used to create a variety of games, including Half-Life: Alyx, Dota 2, and Team Fortress 2. It is known for its versatility and is a popular choice among developers for creating both traditional and VR games. The engine includes a suite of tools and features, such as advanced physics simulations, real-time global illumination, and dynamic AI behavior, which can help developers create immersive and engaging virtual worlds. It is also compatible with multiple platforms, including Windows, Linux, macOS, and various VR headsets. [84]

**Unity Engine** `unity.com/`
Unity is considered the leading cross-platform game engine used by developers to build engaging AR/VR experiences. Unity provides a powerful set of tools and features for building immersive experiences, including support for various platforms, including VR headsets, mobile devices, and smart glasses. Unity's flexible architecture and scripting system make it easy to create complex interactions and animations, while its cross-platform compatibility enables developers to deploy their applications across multiple devices and platforms. There are numerous AR/VR games and applications built using Unity. Some popular examples include Beat Saber, Job Simulator, Superhot VR, Rick and Morty: Virtual Rick-ality, and Pokémon Go. [85]

**Unreal Engine** `unrealengine.com/`
Unreal Engine is a powerful game development engine developed by Epic Games that has been increasingly used for AR/VR applications. It provides developers with a comprehensive set of tools for creating immersive experiences and has been used to build some of the most popular AR/VR applications on the market. The engine includes advanced graphics capabilities and physics simulation, as well as support for popular AR/VR platforms such as Oculus, SteamVR, and Google VR. Unreal Engine's Blueprint visual scripting system also allows developers to quickly prototype and iterate on their ideas without needing extensive coding experience. With its widespread use in the gaming industry, Unreal Engine is a popular choice for creating AR/VR experiences that are both visually stunning and engaging. Unreal Engine has been used to create a number of popular VR games, including "Robo Recall," "Moss," "Tilt Brush," and "I Expect You To Die." [86]

**Windows Mixed Reality** `microsoft.com/en-us/mixed-reality/windows-mixed-reality`
Windows Mixed Reality is a platform developed by Microsoft that combines virtual reality and augmented reality. It uses a combination of sensors, cameras, and software to provide users with an immersive experience. The platform is designed to work with a variety of devices, including head-mounted displays, motion controllers, and mixed reality headsets. Windows Mixed Reality provides users with access to a growing library of immersive content, including games, 360-degree videos, and virtual tours. The platform also allows developers to create their own mixed reality experiences using a variety of tools and technologies. [87]

# Index

*jasminearvr@gmail.com↗